\documentclass[manuscript]{acmart}

\usepackage{color}
\usepackage{float}
\usepackage{subcaption}
\usepackage{xspace} 
\usepackage{graphicx}
\usepackage{multirow}
\usepackage{array}
\usepackage{enumitem}
\usepackage{ctable}
\usepackage{longtable}

\AtBeginDocument{%
  \providecommand\BibTeX{{%
    \normalfont B\kern-0.5em{\scshape i\kern-0.25em b}\kern-0.8em\TeX}}}

\copyrightyear{2022}
\acmYear{2022}
\setcopyright{rightsretained}
\acmConference[ASSETS '22]{The 24th International ACM SIGACCESS Conference on Computers and Accessibility}{October 23--26, 2022}{Athens, Greece}
\acmBooktitle{The 24th International ACM SIGACCESS Conference on Computers and Accessibility (ASSETS '22), October 23--26, 2022, Athens, Greece}\acmDOI{10.1145/3517428.3544814}
\acmISBN{978-1-4503-9258-7/22/10}

\begin{document}

\title{Freedom to Choose: Understanding Input Modality Preferences of People with Upper-body Motor Impairments for Activities of Daily Living}
\renewcommand{\shorttitle}{Understanding Input Modality Preferences of People with Upper-body Motor Impairments}
\author{Franklin Mingzhe Li}
\affiliation{
  \institution{Carnegie Mellon University}
  \city{Pittsburgh}
  \state{Pennsylvania}
  \country{USA}
}
\email{mingzhe2@cs.cmu.edu}

\author{Michael Xieyang Liu}
\affiliation{
  \institution{Carnegie Mellon University}
  \city{Pittsburgh}
  \state{Pennsylvania}
  \country{USA}
}
\email{xieyangl@cs.cmu.edu}

\author{Yang Zhang}
\affiliation{ 
  \institution{University of California, Los Angeles}
  \city{Los Angeles}
  \state{California}
  \country{USA}
}
\email{yangzhang@ucla.edu}

\author{Patrick Carrington}
\affiliation{%
    \institution{Carnegie Mellon University}
    \city{Pittsburgh}
    \state{Pennsylvania}
    \country{USA}
 }
 \email{pcarrington@cmu.edu}

\renewcommand{\shortauthors}{Li et al.}

\begin{abstract}
Many people with upper-body motor impairments encounter challenges while performing Activities of Daily Living (ADLs) and Instrumental Activities of Daily Living (IADLs), such as toileting, grooming, and managing finances, which have impacts on their Quality of Life (QOL). Although existing assistive technologies enable people with upper-body motor impairments to use different input modalities to interact with computing devices independently (e.g., using voice to interact with a computer), many people still require Personal Care Assistants (PCAs) to perform ADLs. Multimodal input has the potential to enable users to perform ADLs without human assistance. 
We conducted 12 semi-structured interviews with people who have upper-body motor impairments to capture their existing practices and challenges of performing ADLs, identify opportunities to expand the input possibilities for assistive devices, and understand user preferences for multimodal interaction during everyday tasks. Finally, we discuss implications for the design and use of multimodal input solutions to support user independence and collaborative experiences when performing daily living tasks. 


\end{abstract}




\maketitle

\section{Introduction}

Activities of Daily Living (ADLs) is a term used to collectively describe fundamental tasks that are required for a person to care for themselves, independently \cite{edemekong2021activities}. They are often broken up into two categories: 1) basic ADLs, such as eating, bathing, and grooming, and 2) Instrumental Activities of Daily Living (IADLs), such as managing finances, shopping, house cleaning, and managing communication. Throughout this paper, we will use ``ADLs'' to refer to tasks in both of these categories. Collectively, these tasks form a base from which society understands a person’s essential independence and begins to evaluate Quality of Life (QOL) \cite{gobbens2018associations}. Since many of these tasks involve interacting with objects and one’s environment using hands and arms, people with upper-body motor impairments can experience difficulties completing these activities \cite{roh2013clinical}. Prior work has shown that people generally adopt one of two solutions to help with ADL tasks: digital assistance through assistive technologies \cite{harada2007voicedraw,zhang2017smartphone}) or human assistance provided by Personal Care Assistants (PCAs) \cite{Personal11:online}. Although there has been prior work to develop assistive technologies that can empower people with disabilities to interact with different computing devices by leveraging alternate input modalities, many users still require human assistance to complete tasks effectively \cite{Personal11:online}.

The reliance on human assistance is due, in part, to a lack of knowledge on how to effectively leverage various input modalities, which have become increasingly common among emerging computing devices. We have not fully explored the potential for multimodal input to support people with upper-body motor impairments in ADL tasks that do not completely rely on computing technology like mobile devices and laptops. Furthermore, we do not fully understand the collaborative role of technology and PCAs in completing ADL tasks. 

To begin addressing these gaps, we conducted an interview study with 12 people with upper-body motor impairments. We first aimed to understand the current practices of people with upper-body motor impairments while performing ADLs, along with the challenges they encounter. We then explored the potential uses of different input modalities during specific daily living tasks. Finally, we highlighted potential applications and implications for the design of multimodal assistive technologies to facilitate user input during ADL tasks. Our research questions are:

\begin{enumerate}[label={RQ\arabic*:}]
  \item What are the current practices and challenges of people with upper-body motor impairments during ADLs?
  \item How and why would various input modalities benefit people with upper-body motor impairments during ADLs?
  \item How can we leverage multimodal input to support people with upper-body motor impairments during ADLs?
\end{enumerate}

In the sections that follow, we first describe upper-body motor impairments and ADLs (Section \ref{ADLs}). We then summarize related literature on both common and emerging input modalities for people with upper-body motor impairments (Section \ref{input modality}) as well as prior research examining multimodal interactions for assistive applications (Section \ref{multimodal}). Next, we describe our semi-structured interview study, the subsequent analysis, 
and our findings. Specifically, we describe practices and challenges of existing approaches toward ADLs (Section \ref{practiceperceptionchallenges}). We interrogate users' preferences for applying individual inputs and combinations of inputs to support specific daily living tasks (Section \ref{findingapplications}). We then describe reasons for individual input modality and multimodal input alternatives (Section \ref{findingimplications}). Finally, we provide design recommendations for future input solutions to support people with upper-body motor impairments during ADLs (Section \ref{Discussion}), including creating multimodal designs that consider collaborative experiences in ADLs, differentiating between interaction with computing devices and systems to support traditional ADLs (e.g., toileting), and consideration for actuation and human-robot interaction during multimodal interactions.


\section{Background and Related Work}
In this section, we first provide background information regarding upper-body motor impairments and the importance of performing ADLs (Section \ref{ADLs}). Next, we describe existing literature that explored different input modalities to assist people with upper-body motor impairments (Section \ref{input modality}). Finally, we show existing research on multimodal input and how it may benefit ways of interactions (Section \ref{multimodal}).

\subsection{Upper-body Motor Impairments and ADLs}
\label{ADLs}
The term \textit{Upper-body motor impairment} usually refers to motor impairments that affect the upper extremities, which are often caused by spinal cord injury, cerebral palsy, muscular dystrophy, etc \cite{WebAIMAr88:online}. People with upper-body motor impairments may have different mobility conditions of their upper limbs, such as fine motor and gross motor impairments. When rehabilitation specialists assess the QOL for people with upper-body motor impairments, they usually evaluate the capability of performing basic and instrumental activities of daily living (i.e., ADLs and IADLs) \cite{gobbens2018associations}. Basic ADLs involve essential tasks such as grooming, cooking, toileting, dressing, and showering, and instrumental ADLs (IADLs) refer to activities that require more planning and thinking, such as using a phone and managing finances \cite{levine2003family}. According to existing research, many people with upper-body motor impairments rely on PCAs for ADLs \cite{Personal11:online} or use inputs to control devices (e.g., voice) to overcome the limitations of upper-body motor impairments \cite{kane2020sense,carrington2014wearables,carrington2016gest,carrington2017chairable}. In order to further support the independence of people with upper-body motor impairments in ADLs and improve the QOL, it is important to first understand how people with upper-body motor impairments perform different ADLs and associated challenges.

\subsection{Input Modalities for People with Upper-body Motor Impairments}
\label{input modality}
Prior research also explored various input modalities for people with upper-body motor impairments to interact with different systems to compensate the upper-body motor impairments \cite{kane2020sense,li2021choose}, such as controlling mobile devices \cite{fan2020eyelid,malu2018exploring,cicek2020designing}, performing text-entry tasks \cite{wobbrock2006trackball,song2010joystick}, interacting with TVs \cite{ungurean2021coping}, using robotic arms \cite{baldi2017design}, and playing games \cite{garcia2013super}. These approaches either rely on readily available devices (e.g., \cite{fan2020eyelid}) or customized technologies (e.g., \cite{ungurean2021coping,baldi2017design}). To interact with these devices, prior research explored various input modalities for people with upper-body motor impairments---touch input \cite{song2010joystick,vatavu2019stroke,carrington2014gest}, hand or arm gestures \cite{ascari2020computer,ascari2020personalized,roy1994gestural}, voice input \cite{harada2007voicedraw,rosenblatt2018vocal,bilmes2005vocal}, eye-based input \cite{fan2020eyelid,fan2021eyelid,zhang2017smartphone,krishna2020eye}, head-movement input \cite{cicek2020designing,rizvi2018simulation}, brain-computer input \cite{dupres2014hybrid,munoz2014application}, facial or mouth-based gestures \cite{wang2018intelligent,grewal2018sip,mougharbel2013comparative}, and biometrics \cite{lazar2017information,nafisah2016evaluating,kim2006practical}. Table \ref{table:inputmodality} summarizes input modalities presented in prior work, from which we took much inspiration in generating our user studies.

\def\arraystretch{1.1}
\begin{table*}[ht]
\centering 

\resizebox{1\textwidth}{!}{%
\begin{tabular}{p{3.7cm}|p{4.8cm}|p{4.8cm}|p{2.6cm}} 

\toprule 
\textbf{Category} & \textbf{Description} & \textbf{Diagnosis of Motor Impairments} & \textbf{Reference} \\ [0.5ex] 
\midrule 
Touch Input & Touch control through joystick, trackball, smartphone, tablet, smartwatch, or customized touchpads. & Spinal cord injuries, spina bifida, orthostatic tremor, cerebral palsy, muscular dystrophy, multiple sclerosis, osteogenesis imperfecta, juvenile rjeumatoid arthritis, radial nerve injury, spastic quadriplegia, hemorrhagic stroke
 & \cite{carrington2014gest}, \cite{malu2018exploring}, \cite{malu2015personalized}, \cite{mott2019cluster}, \cite{naftali2014accessibility}, \cite{song2010joystick}, \cite{vatavu2019stroke}, \cite{wobbrock2006trackball}, \cite{wobbrock2008goal} \\ 
\midrule 

Voice-based Input & Use voice to control different IoT devices, substituting inaccessible input techniques, or performing specific tasks. & Spinal cord injury, cerebral palsy, stroke, spinal dysmorphism & \cite{bilmes2005vocal}, \cite{harada2007voicedraw},  \cite{pradhan2018accessibility}, \cite{rosenblatt2018vocal}  \\ 
\midrule

Eye-based Input & Use eye-gaze position or eyelid gestures to control devices. & Spinal cord injury, amyotrophic lateral sclerosis & \cite{fan2020eyelid}, \cite{zhang2017smartphone} \\ 
\midrule

Head-movement Input & Use head movements and orientations to control a pointer on devices. & Non-specific upper body & \cite{cicek2020designing}, \cite{rizvi2018simulation} \\ 
\midrule

Face-based or Mouth-based Input & Use facial expressions or mouth-based input (e.g., teeth tapping, sip-and-puff) to interact with devices. & Non-specific upper body & \cite{grewal2018sip}, \cite{mougharbel2013comparative}, \cite{wang2018intelligent} \\ 
\midrule

Hand or Arm Gesture & Use personalized hand or arm gestures to interact with computing systems depending on the upper-body mobility. & Cerebral palsy, hydrocephalus, quadriplegia, spastic quadriplegia, static encephalitis, pseudobulbar palsy & \cite{ascari2020computer}, \cite{ascari2020personalized} \\ 
\midrule

Brain-Computer Interface & Use Electroencephalogram (EEG) or Electromyography (EMG) to allow people with upper-body motor impairments to interact with devices or to understand their needs. & Muscular dystrophy, stroke & \cite{dupres2019toward}, \cite{dupres2014hybrid}, \cite{munoz2014application} \\ 
\midrule

Biometric Input & Use biometric information (e.g., fingerprint, voice) for security and privacy purposes. & Cerebral palsy & \cite{kim2006practical}, \cite{nafisah2016evaluating} \\ 
\midrule

Automatic Recognition or Other Input & Use automatic recognition to respond to the user without explicit input (e.g., automatic door opener) or other input modalities. & Non-specific upper body & \cite{bourhis1996mobile} \\[0.5ex] 
\bottomrule
\end{tabular}
}
\caption{Input modalities for people with upper-body motor impairments.}
\Description{This table shows input modalities for people with upper-body motor impairments from existing research. It has four columns and ten rows. The first row presents the headers of each column: Category, Description, Diagnosis of motor impairments, and Reference. For each row, it shows the input method in the first column (left to right), then the description of the input method in the second column, then the diagnosis of motor impairments that were mentioned in the literature that use the specific input method, the final column is the associated references.}

\label{table:inputmodality} 
\end{table*}

\textbf{Touch Input:}
Existing research has explored touch input methods and customization on computing devices to support people who have upper-body motor impairments. For example, Vatavu and Ungurean \cite{vatavu2019stroke} conducted stroke-gesture analyses from a dataset of 9681 gestures collected from 70 participants with motor impairments to outline the research roadmap for accessible gesture input on touchscreens. Similar explorations on touch input have also been conducted using various devices, such as trackballs \cite{wobbrock2008goal,wobbrock2006trackball}, joysticks \cite{song2010joystick}, smartphones \cite{naftali2014accessibility,mott2019cluster}, tablets \cite{vatavu2019stroke}, smartwatches \cite{malu2018exploring}, head-mounted displays \cite{malu2015personalized}, and customized touchpads \cite{carrington2014gest}. 

\textbf{Voice-based Input:}
Voice-based input has been studied to help people with upper-body motor impairments in controlling different IoT devices (e.g., \cite{pradhan2018accessibility}), substituting inaccessible input techniques (e.g., \cite{bilmes2005vocal}), or performing specific tasks (e.g., drawing \cite{harada2007voicedraw}, programming \cite{rosenblatt2018vocal}). For example, Rosenblatt et al. \cite{rosenblatt2018vocal} demonstrated that using vocal input could help people with upper-body motor impairments navigate and edit code in programming.

\textbf{Eye-based Gesture:}
Furthermore, we found that eye-based gestures, such as using eye-gaze fixations \cite{zhang2017smartphone} and eyelid gestures \cite{fan2020eyelid}, could support people with upper-body motor impairments in interacting with digital interfaces. For example, Zhang et al. \cite{zhang2017smartphone} demonstrated decoding eye gestures (e.g., looking up or looking down) into commands that can be used to enable people with upper-body motor impairments to control mobile devices. 

\textbf{Head-movement Input:}
Similar to eye-gaze directions, prior research also leveraged head movements and orientations to control the pointer on a device \cite{cicek2020designing,rizvi2018simulation}. For instance, Cicek et al. \cite{cicek2020designing} proposed a calibration-free head-tracking input mechanism for mobile devices that allows people with upper-body motor impairments to achieve pixel-level pointing precision on small screens. 

\textbf{Face-based or Mouth-based Input:}
Besides using eye-based or head-based input, we found that existing research also explored face-based or mouth-based gestures as an input modality to help people with motor impairments to interact with their devices \cite{wang2018intelligent,grewal2018sip,mougharbel2013comparative,sun2021teethtap}. For example, Wang et al. \cite{wang2018intelligent} introduced the use of facial expressions as controls in games, such as Super Mario Bros., and Grewal et al. \cite{grewal2018sip} showed the approach of using sip-and-puff systems to place commands to a power wheelchair.

\textbf{Hand or Arm Gesture:}
From existing literature, we learned that people with upper-body motor impairments might have various levels and conditions of controlling their upper extremity to interact with their devices (e.g., fine motor, gross motor) \cite{naftali2014accessibility}. Thus, prior research explored various recognition approaches that allow detection of personalized hand or arm gestures for interaction with computing systems based on the abilities of people with upper-body motor impairments \cite{ascari2020computer,ascari2020personalized}. 

\textbf{Brain-Computer Interface:}
In addition, existing research also explored EEG (Electroencephalography) or EMG (Electromyography) approach that allows people with upper-body motor impairments to interact with technologies through brain-computer interfaces (BCI) or to understand their needs \cite{dupres2019toward,munoz2014application,dupres2014hybrid, neuper2003clinical}. For example, Neuper et al. \cite{neuper2003clinical} introduced the approach of EEG-based brain-computer interface to help people with severe motor impairments accomplish tasks like selecting fine-grain letters, where people need to precisely specify the start and end boundary of the selection.

\textbf{Biometric Input:}
As people with upper-body motor impairments are becoming more conscious of privacy and security through daily activities \cite{lazar2017information}, researchers further proposed approaches to leverage biometric information for people with upper-body motor impairments \cite{kim2006practical,nafisah2016evaluating,lewis2021got}, such as fingerprints. For instance, Lewis and Venkatasubramanian \cite{lewis2021got} mentioned the existing approaches of using fingerprint and face recognition to assist people with upper-body motor impairments in completing the authentication process to control their devices.

Overall, we introduced existing research that explored various input modalities to help people with upper-body motor impairments interact with their computing devices for different accomplishments and under certain environments (e.g., \cite{kane2020sense}). Existing work mostly focused on input modalities for computing devices (e.g., computers), which does not fully address the problem of high reliance on PCAs for ADLs (e.g., toileting, dressing). We chose to focus on practices and challenges of leveraging different input modalities in various ADLs. We are also interested in exploring which input modalities are more preferred from the perspectives of people with upper-body motor impairments, such as whether people prefer using voice control or head gestures to tweak the water temperature.

\subsection{Multimodal Input Methods for People with Upper-body Motor Impairments}
\label{multimodal}

Existing research has studied the overall benefits of multimodal input in human activities, such as achieving low-false positive rates \cite{turk2014multimodal} and the accommodation of tasks and context changes \cite{reeves2004guidelines}. More importantly, Reeves et al. \cite{reeves2004guidelines} also implied the potential benefits of multimodal input to adapt to individual differences in mobility conditions (e.g., sensor or motor impairments). Similar to what Reeves et al. \cite{reeves2004guidelines} projected, prior research leveraged multiple input modalities to support people with upper-body motor impairments to interact with their computing devices (e.g., \cite{tomari2012development,wang2018intelligent,dupres2019toward,biswas2012developing,holzinger2006people,wang2019human,liu_crystalline_2022,liu_wigglite_2022}). For example, Wang et al. \cite{wang2018intelligent} combined eye input gestures (e.g., eye movements to the left and right, double blink) with facial expressions (e.g., smile, open and close mouth) to enable people with upper-body motor impairments to provide input for VR games. As another example, Dupres et al. \cite{dupres2019toward} combined hand input with brain-computer interfaces for both better control of applications in daily life (e.g., web browser, video game) and enabling researchers to understand behaviors of people with upper-body motor impairments. Finally, Tomari et al. \cite{tomari2012development} proposed leveraging multimodal input by combining momentary switch and head recognition to control the direction and orientation of smart wheelchairs. 

Although several works leveraged the benefits of combining multiple input modalities to better assist people with upper-body motor impairments \cite{wentzel2022understanding}, there exist gaps between people with upper-body motor impairments in ADLs and how multimodal modalities may benefit the overall input experiences. We are interested in comprehensively investigating multimodal input as opposed to only a few niche interactions and aim to investigate the applications of multimodal input in ADLs. As a result, our research provides guidelines to HCI and Accessibility researchers on designing multimodal input for people with upper-body motor impairments to help with their ADLs.

\section{Method}
\label{methodology}
In this work, we conducted semi-structured interviews with people who have upper-body motor impairments to learn about their existing practices and challenges when performing ADLs, and their preferences for various input modalities. In addition, we investigated opportunities for multimodal input to help people with upper-body motor impairments in ADLs.

\subsection{Participants}
We recruited 12 participants with upper-body motor impairments to participate in our study (Table \ref{table:participants}). Participants were recruited through online platforms (e.g., Reddit, Twitter, Facebook) and snowball sampling. To participate in our study, participants must be 18 years or older, have upper-body motor impairments, have experiences with assistive technologies, and be able to communicate in English. Among the 12 participants we recruited, three of them were female, and nine were male (Table \ref{table:participants}). They had an average age of 31.6 (SD = 8.0). Four participants stated that they had spinal cord injuries, four had cerebral palsy, one had stroke, one had primary lateral sclerosis, one had arthrogryposis multiplex congenita, and one had muscular dystrophy. The study took around 75 to 90 minutes per participant. Participants were compensated with a \$20 Amazon gift card. The recruitment and study procedure was approved by the Institutional Review Board (IRB).

\def\arraystretch{1.15}
\begin{table*}[t]
\centering
\resizebox{1\textwidth}{!}{%
\begin{tabular}{
p{1.6cm}|
p{1.0cm}
p{0.7cm}
p{4.6cm}
p{8cm}
}
    \toprule
    \textbf{Participant} & 
    \textbf{Gender ~~~~} &
    \textbf{Age ~~~~~~~~} & 
    \textbf{Upper-body Motor Impairments} & 
    \textbf{Details}\\
    \midrule
P1 & Female & 41 & Arthrogryposis multiplex congenita & Range of motion and strength of arm and leg are limited. Cannot lift arm or stretch.
 \\
P2 & Female & 23 & Cerebral palsy & Gross motor and fine motor difficulties.
\\
P3 & Male & 35 & Spinal cord injury (C5) & Paralyzed from shoulder down, no finger flexion.
 \\
P4 & Female & 38 & Spinal cord injury (C5) & Some wrist function and no hand dexterity. \\
P5 & Male & 47 & Muscular dystrophy & Strength is limited and bicep is extremely weak, cannot lift the arm without gravity.
 \\
P6 & Male & 27 & Spinal cord injury (C4/C5) & Wrist extension on one side, no finger mobility, no tricep control, no fine motor.\\
P7 & Male & 24 & Cerebral palsy & Paralyzed left arm. \\
P8 & Male & 33 & Spinal cord injury (C5) & Have use of bicep, no triceps, can do to mid way point of the bicep, no sensation further down. No fine motor function on either hand.
 \\
P9 & Male & 22 & Stroke & Right arm cannot go past 45 degrees. \\
P10 & Male & 34 & Cerebral palsy & Difficulty moving wrist, hand, and cannot flex arm.

 \\
P11 & Male & 20 & Cerebral palsy & Floppy limbs due to cerebral palsy. Cannot use the left arm at all. Bicep and tricep functionality are limited.
 \\
P12 & Male & 35 & Primary lateral sclerosis & Have difficulty holding objects and moving around. \\
    \bottomrule
\end{tabular}%
}
\caption{Participants' demographic information.}
\Description{This table shows participants' demographic information. It has five columns in total. The first row presents the headers of each column: Participant, Gender, Age, Upper-body Motor Impairment, and Details. For each row, from left to right, it first shows the participant number, then the gender of that participant, then the age of that participant, next is their self-declared upper-body motor impairments, and the last column is the details of upper-body motor impairments that described by the participants.}
\label{table:participants}
\end{table*}

\subsection{Study Procedure}
\subsubsection{Demographic Background}
In our semi-structured interviews, we first asked the demographic background of our participants (e.g., age, gender, descriptions of upper-body motor impairments). 

\subsubsection{Current Practices and Challenges of ADLs}
We then asked our participants about their current practices of performing certain ADLs \cite{Activiti3:online} (Figure \ref{fig:adls}) and associated challenges across different ADLs (e.g., rely on PCAs, additional). 

\subsubsection{Input Modality Preferences}
Afterwards, we introduced participants with existing input modalities for people with upper-body motor impairments from literature (e.g., head-movement input \cite{cicek2020designing,rizvi2018simulation}, eye-based input \cite{fan2020eyelid,zhang2017smartphone}, brain-computer interface \cite{dupres2019toward,munoz2014application,dupres2014hybrid, neuper2003clinical}) (Table \ref{table:inputmodality}). To ensure participants understood the various input modalities and to reduce bias, we created introduction slides of each input modality by including figures from existing research (e.g., \cite{zhang2017smartphone}) and commercially available products (e.g., \cite{UseDicta29:online}). After introducing each input modality, we confirmed with participants to make sure they understood different input modalities and associated applications. If they still had difficulties understanding the different input modalities, we provided video demonstrations to participants for better understanding. We then asked participants to describe how the different input modalities may benefit their experiences (Table \ref{table:inputmodality}) completing each ADLs (Figure \ref{fig:adls}), and associated reasons.

\subsubsection{Multimodal Input Preferences}
\begin{figure}[t]
    \centering
    \includegraphics[width=0.7\columnwidth]{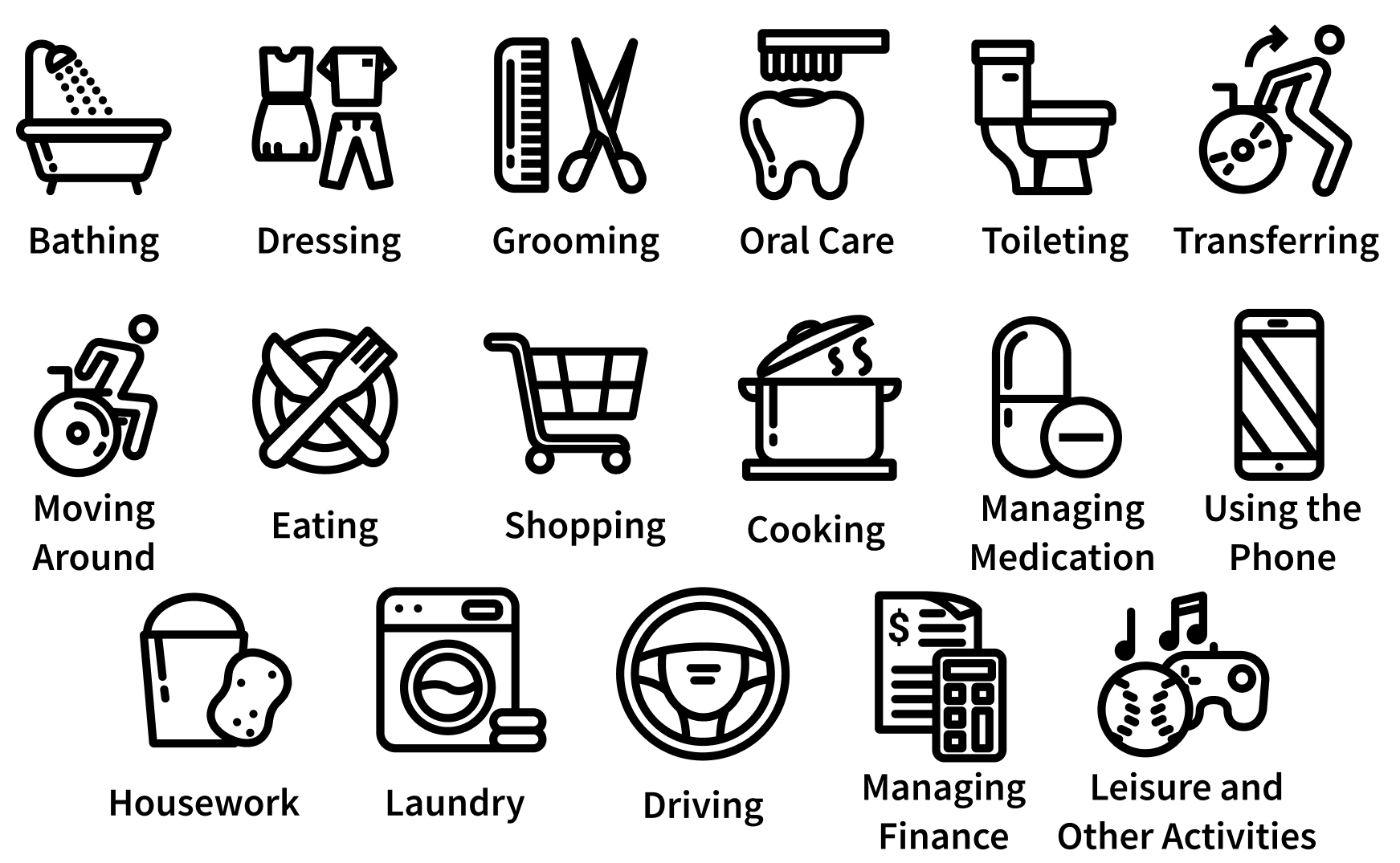}
    \caption{Activities of Daily Living \cite{Activiti3:online}. (Bathing, Dressing, Grooming, Oral Care, Toileting, Transferring, Moving Around, Eating, Shopping, Cooking, Managing Medications, Using the Phone, Housework, Laundry, Driving, Managing Finances, Leisure and Other Activities)}
    \label{fig:adls}
    \Description{This figure contains 17 icons in three rows. The first row has six icons with their meanings underneath each icon, starting from left to right, there are bathing, dressing, grooming, oral care, toileting, and transferring. The second row has six icons with their meanings underneath each icon, starting from left to right, there are moving around, eating, shopping, cooking, managing medication, and using the phone. The third row has five icons with their meanings underneath each icon, starting from left to right, there are housework, laundry, driving, managing finance, and leisure and other activities.}
\end{figure}

Finally, we asked participants for their opinions and preferences on combining different input modalities for each ADL (Figure \ref{fig:adls}) and associated reasons.

\subsection{Data Analysis}
The semi-structured interviews were conducted through Zoom \cite{VideoCon42:online} and all interviews were audio-recorded and transcribed. After the interviews, two researchers independently performed open coding \cite{charmaz2006constructing} on the transcripts. Then the coders met to discuss their codes and resolve any conflicts (e.g., missing codes, disagreement on codes). After the two researchers reached a consensus and consolidated the list of codes, they performed affinity diagramming \cite{hartson2012ux} using a Miro board \cite{AnOnline70:online} to cluster the codes and identify emergent themes. 


\section{Findings: Current Practices and Challenges of ADLs by People with Upper-body Motor Impairments}
\label{practiceperceptionchallenges}
In this section, we first show existing practices of ADLs by our participants (Section \ref{currentpractice}). We then present associated challenges in ADLs from people with upper-body motor impairments (Section \ref{perceptionschallenges}). 
\begin{table*}[ht]
\centering 

\resizebox{1.00\textwidth}{!}{%
\begin{tabular}{
>{\raggedright}p{19mm}|
>{\raggedright}p{18mm}|
>{\raggedright}p{62mm}|
>{\raggedright}p{18mm}|
p{54mm}
} 

\toprule 
\textbf{Activities of Daily Living} & 
\textbf{Number of Participants Required PCAs} & 
\textbf{Sample Tasks through Personal Assistance} & 
\textbf{Number of Participants Use Technology} & 
\textbf{Sample Tasks through Technology} \\ [0.5ex] 
\midrule 
Bathing & 10 & Transfer between wheelchairs and shower bench, control water temperature, dry after a shower, turn on/off the water & 2 & Turn on/off the water, control water temperature \\ 
\midrule 

Dressing & 11 & Help with zipping, help with small buttons, feel the clothes to make sure it is flat or smooth & 1 & Use tablet to select clothes \\ 
\midrule 

Grooming & 8 & Comb the hair, trim nails, shaving & 6 & Customized electric hair removing reduce the required range of motion, automatic hand-wave dispenser\\ 
\midrule 

Oral Care & 9 & Turn on/off electric toothbrush, cleaning & 11 & Electric toothbrush, electric flosser \\ 
\midrule 
Toileting & 12 & Transfer, flushing, cleaning & 3 & Automatic height adjusting hydraulics, voice control flushing systems \\ 
\midrule 
Transferring & 9 & Transfer from wheelchair to bed & 8 & Seat elevation, joystick and remote control to change into a comfortable position \\ 
\midrule 
Moving Around & 3 & Open/close doors & 9 & Different input modalities to control the wheelchair moving direction and speed \\ 
\midrule 
Eating & 11 & Choose the food on the plate, lift utensils, cut the food & 3 & Omnibot to scoop and feed the person, spoon with accelerometer that could level itself \\ 
\midrule 
Shopping & 9 & Overhead reach of products, put food in bags, carry food & 7 & Online shopping through accessible smart devices \\ 
\midrule 
Cooking & 11 & Slicing, getting hot plates off the stove, everyday cooking & 4 & Voice control microwave, electric cooker, automatic peeler and cutter \\ 
\midrule 
Managing Medication & 8 & Pill refill, picking up drugs & 6 & Pill organizer, reminders \\ 
\midrule 
Use the Phone & 1 & Plug in the charger & 10 & Smaller phone for touch with less range of motion, use voice to reduce the need of motion  \\ 
\midrule 
Housework & 10 & Mobbing the floor, dish washing, vaccuming & 7 & Robot sweeper, roomba, voice-based light control, TV control \\ 
\midrule 
Laundry & 11 & Folding clothes, general laundry & 6 & Customized dials and buttons to control the washer, voice-assisted camera systems to place commands and instructions \\ 
\midrule 
Driving & 9 & Change settings that are hard to reach, general driving & 5 & Hand control of A/C, touchscreen for radio,  \\ 
\midrule 
Managing Finances & 4 & Accessing mailbox, write the information on a check, count cash & 11 & Keep records on computers, camera to deposit checks, online banking \\ 
\midrule 
Leisure and other activities & 1 & Horse riding & 8 & Play video games, walk dog, control TV \\[0.5ex] 
\bottomrule

\end{tabular}
}
\caption{The number of participants that leveraged technologies or PCAs for each ADL and associated sample tasks.} 
\Description{This table shows the number of participants that leveraged technologies or PCAs for each ADL and associated sample tasks. This table has five columns. The first row presents the header of this table: Activities of Daily Living, Number of Participants Required PCAs, Sample Tasks through Personal Assistance, Number of Participants use Technology, and Sample Tasks through Technology.}
\label{table:ADLPractice} 
\vspace{-5mm}
\end{table*}

\subsection{Practice of ADLs by People with Upper-body Motor Impairments}
\label{currentpractice}
In terms of current practices of ADLs, we learned that participants \textbf{rely heavily on Personal Care Assistant (PCA) or their family members} to help with different ADLs (e.g., dressing, bathing, toileting, driving) (Table \ref{table:ADLPractice}). Based on the responses to each activity of daily living and statistical analysis, we found that about 67.2\% (SD = 29.7\%) of participants require PCAs for each ADL on average (Table \ref{table:ADLPractice}). P3 commented on this situation:

\begin{quote}
    ``...Many of the activities still require my personal assistant for help...tasks like doing my laundry, toileting, and shopping...I have a very limited range of motion, and current technologies are not there yet to support me living independently...''
\end{quote}

We also found that there is a \textbf{high disparity between ADLs that involves computing devices (e.g., managing finances) and basic ADLs in the reliance on PCAs (e.g., toileting, dressing)}. Only 10\% of our participants require other people for assistance with using the phone and leisure activities (Table \ref{table:ADLPractice}). However, for activities like toileting, dressing, and cooking, over 90\% of our participants still require other people for assistance (Table \ref{table:ADLPractice}). P5 explained this:

\begin{quote}
    ``...Traditional daily activities or essential activities are usually restricted based on my motion capability and also the inaccessible environment. One example is that my washer and dryer are in the basement, which forced me to use PCAs for assistance...''
\end{quote}

Among all technologies that are being used to help people with upper-body motor impairments in ADLs, we found that the majority of \textbf{existing input systems for traditional ADLs require additional effort for installation, adjustment, and modification supports for individuals with upper extremity capabilities}. For example, P12 explained how his multimodal shower control system was installed based on the mobility of his upper extremity:

\begin{quote}
    ``...My cousin helped me to use my voice to control the temperature and added a big button on the left side about my shower seat height to allow me to use my less affected left hand to turn on or off the water...He had to seal all wires, the button, and the microphone...He also needs to program and map my preferred way of controlling the water...''
\end{quote}

Furthermore, we realized that some of our participants \textbf{leverage both technology and PCAs to collaboratively accomplish certain ADLs}. For example, P11 explained how he could use the electric toothbrush by himself, but needed someone to put toothpaste on the brush and press the on/off button:

\begin{quote}
    ``...I have limited motion control of my arm, which made me hard to press the on/off button on the electric toothbrush, and it is a tiresome process to put toothpaste on the brush as well. But I could use the limited motion control to brush my teeth after the power is on. So I usually have my personal assistant add toothpaste for me and turn on the toothbrush, so she can focus on other things while I am brushing my teeth...''
\end{quote}

\subsection{Challenges with ADLs by People with Upper-body Motor Impairments}
\label{perceptionschallenges}

According to what we showed above, we found that there exist unique practices for people with upper-body motor impairments across different ADLs. Thus, we further show challenges of different existing approaches toward ADLs and associated challenges. From the interview, we learned that participants in our study rely heavily on PCAs or their family members to help with different ADLs (e.g., dressing, bathing, toileting, driving). Despite the usefulness and necessity of having PCAs for ADLs, they also mentioned limitations and concerns of having other people assist with ADLs. For example, P2 mentioned that always relying on PCA for help with ADLs may affect her \textbf{choice of time to do certain ADLs}. P2 further elaborated on this:

\begin{quote}
    ``...My PCA helped me with basically everything in my daily activities. I appreciate everything. However, this forced me to only be able to do certain activities while my PCA was around. For example, toileting, bathing, and dressing. It would be impossible if I wanted to go out and visit my friend at a certain time without having my PCA put my clothes on...''
\end{quote}

Furthermore, P6 complained about the \textbf{financial burden of hiring personal assistants}: 

\begin{quote}
    ``...I am already low income, and hiring personal assistants is a huge part of my monthly cost. If I can do some tasks myself with my voice or eye, it would reduce the number of times that I need personal assistants per month...''
\end{quote}

We also learned that our participants have \textbf{concerns about privacy} while relying on other people for assistance in certain ADLs, such as dressing, toileting, and bathing. P2 further commented:

\begin{quote}
    ``...Although having a personal assistant is the only way to help me with some activities like toileting and bathing, sometimes I feel embarrassed when they see me using the bathroom or having a shower. Especially when a new PCA comes...''
\end{quote}

Moreover, we found that there also exists \textbf{communication barriers} between PCAs and people with upper-body motor impairments, such as language barriers (P2) and difficulties in verbally describing the detailed instructions if the person with upper-body motor impairments cannot be physically in front of certain appliances (P2, P4). P2 further commented on this: 

\begin{quote}
    ``...If a PCA doesn't know English well, it's hard for me to communicate with the PCA regarding my needs, especially for doing the laundry where my washer is in the basement that is not accessible...like putting items that shrink separately, and putting bras in the bra bags...''
\end{quote}

Besides the concerns and challenges while working with PCAs, five of our participants also expressed concerns about the inability to \textbf{contribute to housework when their family members are busy}. P8 commented on his situation:

\begin{quote}
    ``...I live with my family. My sister and parents usually help with various daily activities, such as doing laundry and bathing. I know they are very busy with other stuff too, and I really want to help them sometimes. For example, I would like to cook them some meal before they come back home, but existing technology does not support me to do so...''
\end{quote}

In terms of their current challenges of existing technologies, all of the participants expressed limited input options they have obtained and the willingness to try new input modalities. Specifically, four participants mentioned that existing input systems mostly rely on a fixed single input modality for a particular task. They mentioned the \textbf{concerns of convenience, effort, and reachability for a fixed input modality} (P3, P6, P7, P9). For example, P7 explained the limitation of using hand waving to control the A/C of his car when he does not sit on the drivers' seat:

\begin{quote}
    ``...I had my car customized by allowing me to use hand-waving gestures to control my A/C while driving. However, this only works when I am in the driver's seat. It becomes unreachable for me while I am sitting in the passenger's seat...''
\end{quote}

Beyond the fixed input modality, three participants mentioned concerns about the \textbf{fixed mapping after installation} between certain input and specific tasks, and it is almost impossible to modify the mapping by themselves. P5 commented on her customized dial pads to control the washer: 

\begin{quote}
    ``...I had a technician install this dial pad to control my washer a couple of years ago. At that time, I was able to reach the top level of buttons to control the water level and the heat, but because of the muscle decline, reaching the top row became hard for me last year, and I did not know how to reprogram the pads...''
\end{quote}

Compared with having personal assistance for ADLs, the use of technology could potentially address concerns from participants regarding privacy concerns, time and effort requirements from PCAs, and maintain self-confidence in social interactions. Overall, we showed the current limitations of existing technologies in ADLs and found that all of our participants showed strong needs and preferences for being able to use various input modalities in ADLs towards independence. Therefore, it is important to uncover their preferences among different input modalities in ADLs and how these input modalities may help people with upper-body motor impairments in accomplishing ADLs by maintaining independence.

\section{Findings: Applications of Individual Input Modality and Multimodal Input in ADLs}
In this section, we first present preferred applications of individual input modality in ADLs for people with upper-body motor impairments (Section \ref{individualpreference}). We then show participants' preferences on applications of different combinations of input modalities (Section \ref{Applications of Multiple Modality Inputs}).
\label{findingapplications}

\subsection{Applications of Individual Input Modality}
\label{individualpreference}

\label{Applications of Certain Input Modality on ADLs}
In our interview, we asked participants whether they would like to change their existing ways of doing ADLs and what input modalities (Table \ref{table:inputmodality}) they would prefer to use for specific ADLs (Fig. \ref{fig:adls}). On average, almost all (11/12) participants mentioned that they would like to change their existing method for ADLs (Table \ref{table:ADLPractice}). Among the participants who had already been using technologies in ADLs, nearly all of them would like to change the technology involved by having new input modalities. Only one participant did not want to change the technology involved for toileting, transferring, housework, or leisure and other activities. For participants who do not use technologies in certain ADLs, we found that all of them prefer having new input modalities that enable them to accomplish these ADLs independently or collaboratively, especially for the ADLs that they may fully rely on PCA or family members, such as dressing, bathing, toileting, and eating.

While we examined each ADL, it became clear that certain input types are favored over others. Across all tasks, we found that \textbf{touch} (e.g., joysticks, touchscreens) and \textbf{voice inputs} are highly desirable compared with others (Table \ref{table:inputpreference}). On average, seven participants prefer using touch or voice input as an input modality to substitute their existing ways of handling ADLs. Specifically, we found that our participants mostly prefer using touch input for bathing (10), toileting (8), and cooking (8). Furthermore, our participants would also like to use voice input for cooking (10) and driving (10) (Table \ref{table:inputpreference}). For example, P8 mentioned that he would prefer using touch input rather than the existing shower knob:

\begin{quote}
    ``...I usually have my dad or sister help with the bathing process because I cannot rotate the shower knob due to the lack of control with my fingers and hands. However, I can `tap' with my palm, I can definitely use the touch interface to turn on/off the water or control the water temperature...''
\end{quote}

Beyond touch or voice input, we learned that \textbf{hand-only gesture} (e.g., waving hands) is the third preferred input method by our participants, which has about three participants prefer using hand-only gestures as an efficient input method for each ADLs. Specifically, we found five participants preferred hand-only gestures for cooking, and four participants chose hand-only gestures for grooming, toileting, and housework (Table \ref{table:inputpreference}). P3 commented on his preferences of waving his hand to flush the toilet or control cooking appliances:

\begin{quote}
    ``...I have spinal cord injuries, and it is hard for me to push or press buttons with my fingers. I think waving hands is a very compelling way for me to control or interact with my devices. Especially for toileting, I always have trouble pressing the button to flush the toilet after use, it would be better to wave my hand after use to flush it...For cooking, I also have trouble using the knob to control the time and temperature. Thus, using waving to control a scrolling panel would be a great option for me...''
\end{quote}

\textbf{Eye-based input} (e.g., eye gaze, eyelid gestures), \textbf{head-movement input}, and \textbf{brain-computer input} all have the same number of selections (2) by participants who prefer to use these methods for all ADLs. By further analyzing the data quantitatively, we found that participants highly prefer eye-based input (6) and head-movement input (5) for shopping (Table \ref{table:inputpreference}). P10 explained her existing way of shopping and why he prefers eye-based input: 

\begin{quote}
    ``I do not often go shopping physically, I usually just use online platforms such as Amazon, for most of the things. The main barrier is that I cannot buy things at a store by myself. If it is possible, I want to use eye-based input to dwell at a product, and a robotic arm can grab that product for me...''
\end{quote}

Finally, we learned that our participants prefer to use \textbf{facial gestures}, \textbf{biometric input}, and \textbf{automatic recognition or other input} only for limited tasks, which ends up being the least selected input modalities among all ADLs (1). For instance, we found participants mostly prefer using biometric input for managing finances, because biometric information could be used for authentications. In terms of automatic recognition or other input, seven participants prefer it when in motion, P6 commented on the difficulty of moving physical obstacles (e.g., door):

\begin{quote}
    ``Opening my door is the hardest thing ever, I am always in my wheelchair, and I do not have much control of my upper body, which is not sufficient to open my front door, which forced me to ask my family members to open the door for me all the time. This is why I want an automatic door that can open when I approach...''
\end{quote}

\def\arraystretch{1.15}
\begin{table*}[t]
\centering
\resizebox{1\textwidth}{!}{%
\begin{tabular}{
p{3.6cm}|
p{1.8cm}|
p{1.8cm}|
p{1.7cm}|
p{1.6cm}|
p{2.4cm}|
p{2.3cm}|
p{2.2cm}|
p{1.6cm}|
p{0.9cm}
}
\toprule
    & \textbf{Touch Input} 
    & \textbf{Hand-only Gestures} 
    & \textbf{Voice Input}
    & \textbf{Eye-based Input} 
    & \textbf{Head-movement Input}
    & \textbf{Brain-computer Input}
    & \textbf{Facial Gestures}
    & \textbf{Biometrics}
    & \textbf{Others} \\\midrule
Bathing                      & 10                                                              & 3                                                           & 9                               & 2                                                                       & 3                                       & 3                                        & 5                                                                & 0                                                                    & 0                                                                                            \\
Dressing                     & 4                                                               & 3                                                           & 8                               & 2                                                                       & 2                                       & 0                                        & 2                                                                & 0                                                                    & 1                                                                                            \\
Grooming                     & 7                                                               & 4                                                           & 6                               & 3                                                                       & 3                                       & 1                                        & 3                                                                & 1                                                                    & 2                                                                                            \\
Oral Care                    & 5                                                               & 2                                                           & 2                               & 3                                                                       & 3                                       & 2                                        & 4                                                                & 2                                                                    & 1                                                                                            \\
Toileting                    & 8                                                               & 4                                                           & 6                               & 0                                                                       & 3                                       & 2                                        & 3                                                                & 0                                                                    & 2                                                                                            \\
Transferring                 & 5                                                               & 2                                                           & 5                               & 1                                                                       & 2                                       & 1                                        & 1                                                                & 0                                                                    & 1                                                                                            \\
Moving Around                & 6                                                               & 2                                                           & 9                               & 1                                                                       & 3                                       & 2                                        & 0                                                                & 6                                                                    & 7                                                                                            \\
Eating                       & 7                                                               & 2                                                           & 6                               & 4                                                                       & 2                                       & 2                                        & 0                                                                & 0                                                                    & 1                                                                                            \\
Shopping                     & 8                                                               & 3                                                           & 7                               & 6                                                                       & 5                                       & 3                                        & 0                                                                & 0                                                                    & 0                                                                                            \\
Cooking                      & 8                                                               & 5                                                           & 10                              & 2                                                                       & 1                                       & 1                                        & 0                                                                & 0                                                                    & 1                                                                                            \\
Managing Medications         & 7                                                               & 1                                                           & 7                               & 2                                                                       & 0                                       & 3                                        & 0                                                                & 1                                                                    & 2                                                                                            \\
Uses the Phone               & 7                                                               & 3                                                           & 8                               & 5                                                                       & 3                                       & 2                                        & 1                                                                & 4                                                                    & 0                                                                                            \\
Housework                    & 6                                                               & 4                                                           & 9                               & 3                                                                       & 1                                       & 2                                        & 0                                                                & 0                                                                    & 0                                                                                            \\
Laundry                      & 8                                                               & 2                                                           & 6                               & 0                                                                       & 0                                       & 2                                        & 0                                                                & 1                                                                    & 1                                                                                            \\
Driving                      & 8                                                               & 2                                                           & 10                              & 2                                                                       & 3                                       & 1                                        & 1                                                                & 2                                                                    & 0                                                                                            \\
Managing Finances            & 6                                                               & 0                                                           & 5                               & 2                                                                       & 1                                       & 0                                        & 0                                                                & 8                                                                    & 1                                                                                            \\
Leisure and Other Activities & 8                                                               & 2                                                           & 6                               & 3                                                                       & 3                                       & 4                                        & 0                                                                & 0                                                                    & 0 \\
\bottomrule
\end{tabular}
}
\caption{Input modality preferences for specific ADLs among all 12 participants.}
\Description{This table presents input modality preferences for specific ADLs among all 12 participants. There are ten columns in total. The first column, from left to right, shows different ADLs. From the second column to the right most column shows different input modalities. Each cell presents a number that represents the number of participants who prefer to use a specific input modality for the specific ADL.}
\label{table:inputpreference}
\end{table*}

\subsection{Applications of Multiple Modality Inputs}
\label{Applications of Multiple Modality Inputs}
We further identify participants' preferences for applications of different combinations of input modalities. The combination means participants either prefer using several input methods for redundancy purposes or use these input methods to accomplish different sections of a specific task. For example, people may prefer using both touch input and voice-based input for switching the water temperature during showering. On the other side, people may have eye-based input to target an object and use voice-based input to open or close the object. 

In our paper, if a certain combination of input modalities got selected more than five times by our participants, we define it as a popular combination. Based on participants' responses, we found that there is a majority popular combination (selection > 5 among all participants) between touch and voice input for the majority of ADLs, which include \textbf{bathing, grooming, toileting, eating, cooking, managing medications, use the phone, housework, laundry, driving, and leisure and other activities} (Table \ref{table:inputpreference}). We found \textbf{moving around} has the most number of input modalities in popular combinations (selection > 5 among all participants), which includes touch input, voice-based input, biometric input, and automatic recognition or other input. From what we uncovered in the interviewees' responses, we found \textbf{moving around} usually involves more complex environments. Having multiple input modalities would allow people to easier accommodate the complexity of such interactions. We also realized people prefer combining touch input, voice-based input, and eye-based input (selection > 5 among all participants) for \textbf{shopping} specifically (Table \ref{table:inputpreference}). Furthermore, we found that \textbf{managing finances} has a popular combination (selection > 5 among all participants) between touch input and biometric input. 

\section{Findings: Choosing Input Alternatives for Performing ADLs}
\label{findingimplications}
We showed preferred applications of both individual input modality and multimodal input in the previous section. In this section, we first show the reasons for choosing individual input modality alternatives in ADLs (Section \ref{Comparison between Each Input Modality in ADLs}). We then present the reasons for choosing multimodal input alternatives (Section \ref{multimodalbenefits}).

\subsection{Reasons for Individual Input Modality Alternatives}
\label{Comparison between Each Input Modality in ADLs}
In the interview, we asked our participants why they would prefer one input modality over the other based on their responses to input preferences that we showed in the prior section. We uncovered five main factors (i.e., usability, efficiency, consequences, personalization, and context) that may affect how people with upper-body motor impairments choose a specific input modality, and we present each of them in detail.

\textit{\underline{\textbf{Usability:}}} We learned that participants highly value \textbf{reliability and confidence} while choosing a specific input modality (P6, P8, P10, P11). For example, P10 mentioned that he prefers touch input over voice-based input for some ADLs because he knows what the outcome of the touch interaction will be, and it makes him become more confident with providing reliable output. P11 further explained the involvement of biometric information in driving to ensure reliability and reduce false positives:

\begin{quote}
    ``...I can only use my right side of the body to drive. This makes driving a hard task for me because there are so many different functions, and I do not have control of my left hand or foot, which causes the huge concern of accidentally touching and safety issues. That is why I prefer to use fingerprint as a verification approach to make sure I do not accidentally touch somewhere and prevent false activation...''
\end{quote}

Another important factor for choosing an input modality is \textbf{input precision}, which indicates how precise the user can place the command with certain input modalities (P4, P8, P9). For instance, P9 mentioned that he prefers using touch input for some activities because it can make precise commands. P4 further commented on why she likes using voice or joystick compared with hand-only gestures for certain ADLs:

\begin{quote}
    ``...For tasks that need you to set up a specific time or temperature, using my voice or joystick would be easier. You can simply say, `turn the water temperature to 39 degrees,' but it would be very difficult if you use hand gestures to indicate such information...''
\end{quote}

Furthermore, we found that \textbf{learnability and intuitiveness} affect how people with upper-body motor impairments choose to use a certain input modality for ADLs (P4, P9, P10). This means that they prefer the input modality have a lower learning curve (P10). P9 further compared voice-based input with hand gestures in new skill training efforts:

\begin{quote}
    ``...Voice is more straightforward to me. If I need to place commands for my kitchen appliances, I have to learn different hand gestures that map to various commands. However, using voice would not require me to spend time to learn and remember certain commands...''
\end{quote}

Similar to learnability, we also uncovered that our participants highly value the importance of \textbf{adaptability and compatibility}, which means using a single input modality on a specific system should be able to interact with different devices for various purposes (P8, P9, P10, P12). P10 commented on how he likes to use the same touchscreen system to both control the kitchen appliances at home and help with navigation in grocery stores:

\begin{quote}
    ``...I hate switching input devices for different purposes due to the difficulty of switching the hardware and adapting new methods based on mobility conditions. Therefore, I would like to control or place commands from a single unit with a single method, such as having my touchscreen connect to my instant pot while cooking and also accessibility maps of the grocery store for navigation. This compatibility will reduce the effort of changing devices for different purposes...''
\end{quote}

\underline{\textit{\textbf{Efficiency:}}} From the interview, we found that our participants mentioned the necessity of ensuring efficiency for a specific input modality. First of all, \textbf{response time} is a key factor for certain input modalities to be used for some specific ADLs (e.g., temperature change in showering) (P4, P6, P7, P8, P11). P6 further elaborated on the importance of quickly changing temperature while taking a shower for people with spinal cord injury:

\begin{quote}
    ``People with spinal cord injuries care about skin temperature a lot because of the lack of sensation. I prefer hand gestures over voice while changing the water temperature, because using voice can take a long time, including activation, placing specific commands, and waiting for the responses from the system. I might get burned with hot water by that time.''
\end{quote}

Moreover, some people with upper-body motor impairments (e.g., spinal cord injury, cerebral palsy) have a hard time making movements both for the upper limbs and also their lower limbs as well. Therefore, they mentioned the importance of \textbf{less physical effort} (P3, P4, P5, P6, P9). P3 explained the necessity of less movement required for certain ADLs:

\begin{quote}
    ``...I have a C5 level of spinal cord injury and have to stay in my wheelchair all the time. Currently, most of the devices now require me to physically move close to them and touch them to place commands. This situation is frustrated because some spaces are too narrow for me to move around and take turns. My bathroom is a good example, that is why I would prefer to use eye-based input or brain-computer input as an input compared with touch input for these interactions that requires me to  move around...''
\end{quote}

Besides requiring physical effort, we learned that our participants also mentioned the preferences of \textbf{simpler level of interactions} according to the mobility conditions when performing the input (P1, P4, P8). P8 commented on this:

\begin{quote}
    ``...Compared with different hand gestures, I like simple input methods, just like eye blinking or voice. I might not be able to perform certain gestures, and it will end up with poor performance...''
\end{quote}

\underline{\textit{\textbf{Consequences:}}} Our participants mentioned the concerns of various consequences after using certain input modalities, which might change their minds about continuing with that specific input method. First, both P6 and P9 commented on the importance of \textbf{mess prevention} while using certain input modalities (e.g., touch) in ADLs (e.g., bathing, cooking). P6 explained his concerns about using touch interfaces, but rather hand gestures or voice while having a shower:

\begin{quote}
    ``...I like using touch screens because they can provide me with accurate output. However, it can also be problematic when the touch interfaces already have water on them, or my hand has shampoos. For this situation, I will just use hand gestures or voice input in a better way without creating a mess in the bathroom...''
\end{quote}

Next, we also learned that \textbf{safety} is an important factor for people with upper-body motor impairments when choosing input modalities for ADLs (P1, P3). For example, P3 mentioned the importance of having hands-free interactions because he only has control of the arm and no sensations of the hand or finger. P3 further commented on this:

\begin{quote}
    ``I cannot tell if my finger gets burned or not due to the loss of sensation from spinal cord injury, so I would prefer not using touch input while cooking or boiling water.''
\end{quote}

Beyond safety concerns, our participants also mentioned the importance of \textbf{confidentiality and security} for input modalities (P1, P6, P8, P11). P8 mentioned the concerns of confidentiality while checking out as a person with upper-body motor impairments and why he prefers biometric input for payment:

\begin{quote}
    ``...I often have a hard time pulling out my credit card and paying at the cashier. Not just that, due to the difficulty of moving my upper body, it can take a long time to finish the payment process, and other people behind me may see my personal belongings in my wallet, which can be dangerous to me. That is why I think biometric input could reduce the consequences of breaching my personal identities...''
\end{quote}

\underline{\textit{\textbf{Personalization:}}} Participants in our study had different motor impairments affecting their bodies in different ways, including their upper and lower limbs. To account for these differences, personalized input modalities are preferred when selecting a specific input method. For example, P7 and P9 mentioned the importance of taking \textbf{physical ability and mobility} into consideration. P7 further explained:

\begin{quote}
    ``...The limited mobility of my hand affects my choices of input methods, which means I would prefer the input method that can take my personal situations into consideration. This is the reason why I prefer eye movement over touch...''
\end{quote}

Furthermore, we learned that \textbf{familiarity} is another factor in choosing personalized input modalities (P1, P5, P9). We found that our participants preferred certain input modalities because they had already used them for other ADLs. P1 further elaborated on her existing experiences with the joystick, which made her prefer touch input: 

\begin{quote}
    ``...Touch input would be good for me to do other tasks if the technology is available. Because I am already familiar with different ways of using joystick, touchscreen, and trackball. I have been driving with joystick for many years...''
\end{quote}

Similar to what we have mentioned previously (Section \ref{perceptionschallenges}), our participants prefer using input modalities that can enable \textbf{independence} for specific ADLs based on individual circumstances (P2, P4). P2 further commented:

\begin{quote}
    ``...I prefer input methods that can ensure independence when I am taking a shower, eating, toileting or dressing. More specifically, I do not have to wait for my PCA all the time, I can then eat whenever I want...''
\end{quote}

\underline{\textit{\textbf{Context:}}} We identified that the context of specific environments also affects how people perceive different input modalities. For instance, six participants mentioned that \textbf{environmental influences} affect how they choose a specific input method. The major consideration was using voice as input while the surrounding environment is noisy, which may affect the accuracy. P3 commented on his preferences during gaming:

\begin{quote}
    ``...My body limits me from using touch input for aiming in playing FPS games...I do not like voice input when playing games because either the game sound is noisy and the recognition system cannot detect my voice correctly, or I have to use my voice to communicate with my teammates from the game. Therefore, I would go a lot with eye-based input and simple touch commands as a confirmation for shooting...''
\end{quote}

In addition, P6 further mentioned the importance of \textbf{reachability} in a space by using certain input modalities, which includes considerations of both the space and scalability:

\begin{quote}
    ``...Some input methods require certain space for interactions, such as eye tracking, head tracking, and brain-computer interfaces. I do not have a huge space at home. This brings the concerns of whether I should use a certain input for daily tasks...''
\end{quote}

Finally, participants also expressed concerns regarding the \textbf{social acceptability} of using certain input modalities during interactions with or around other people (P5, P6, P8, P9). Three of them explicitly mentioned that using voice input in a store or during shopping can be embarrassing and attract unwanted attention. P6 further explained this:

\begin{quote}
    ``...It is less socially acceptable to use voice input at a store, and I feel it is weird to have someone see me yelling at my wheelchair or a specific product in a store. That is why I would prefer to use eye-based or touch interfaces that have gentle reactions from other people...''
\end{quote}

\subsection{Reasons for Multimodal Input Alternatives}
\label{multimodalbenefits}
We asked them about their preferences for combining various input modalities together. We illustrate reasons for two main uses of multiple modality input: multi-modality for input redundancy (Section \ref{Multimodal for Input Redundancy}) and multi-modality for input variability (Section \ref{Varied Inputs throughout Task Completion}).

\subsubsection{Multi-modality for Input Redundancy}
\label{Multimodal for Input Redundancy}
Based on our participants' responses about why they prefer having multimodal input for redundancy, we uncovered four main factors (i.e., reliability, convenience, customization, appropriateness) that affect how and why our participants choose multiple input modalities for the same purpose vs. just having a single input modality. 

\underline{\textit{\textbf{Reliability:}}} we learned that our participants prefer having redundant input methods for the same purposes to make sure they can accomplish tasks in a reliable way. The first reason our participants mentioned is that \textbf{having a backup input modality} could enable people with upper-body motor impairments to accomplish certain tasks disregarding the physical and mobility limitations (P2, P4, P5, P8). P5 further commented on this:

\begin{quote}
    ``...I have muscular dystrophy, I continuously lose my muscles every day. On the one hand, having multiple input methods for the same purpose could allow me to accommodate my muscle decline as time passes. On the other hand, having two or three options could allow me to finish the tasks with the most suitable way of input based on my physical situation. For example, I could use voice to turn on the water remotely if I want to have a bath while I am in my wheelchair, this just gives me more options for specific tasks...''
\end{quote}

Another reason is that having multiple inputs for redundancy could \textbf{ensure task completion through environmental changes} (P1, P2, P4). We mentioned this in Section \ref{Comparison between Each Input Modality in ADLs} as the reason why they prefer certain input modalities. P4 further commented on having redundant input while taking a shower:

\begin{quote}
    ``...The benefit of having multiple inputs could benefit the showering experiences for sure. Because the environment could be very noisy at that time, if I only used voice, it might not work nicely. That is why having touch input as an alternative is a great option...''
\end{quote}

Beyond always having a backup and ensuring task completion through environmental changes, we found that our participants consider redundant inputs as having \textbf{better accuracy and precision} for various tasks (P4, P10). P4 further commented on this:

\begin{quote}
    ``...Different input methods usually have different precision and accuracy. Take showering as an example, I would like to have both voice control and joystick as the input methods to control the water. However, a joystick is more in control and more precise than just using voice...''
\end{quote}

\underline{\textit{\textbf{Convenience:}}} we found that five participants claimed that they like redundant input mainly for convenience. First, P4, P6, and P12 commented on the importance of maintaining \textbf{multi-purposing} during complex situations, which allows them to choose their preferred input method freely. For example, P4 mentioned her experiences of having two or more things at a time: 

\begin{quote}
    ``...I sometimes have multiple things happening at the same time, such as I may have to answer the phone while I am driving through touch. Therefore, having multiple input methods for different tasks offers me great flexibility in daily activities...''
\end{quote}

Moreover, we found redundant input offers better \textbf{reachability} for participants with upper-body motor impairments due to mobility issues. It allows people to control their devices at a certain distance with comfort. P1 commented on her preference of combining the joystick with the voice to control the seat elevation while transferring:

\begin{quote}
    ``...I use seat elevation to do some of the transfers, such as transferring from my wheelchair to my bed. After I raise my wheelchair, I need someone else to put the wheelchair back down for transferring back because I can only control the seat elevation through the joystick on my wheelchair. If I can use voice to control my wheelchair, I won't need someone else for assistance in transferring...''
\end{quote}

\underline{\textit{\textbf{Customization:}}} we learned that our participants preferred having redundant input for customized input opportunities to \textbf{compromise their own mobility conditions}. P6 commented on her situation of only being able to perform a portion of gestures through her hands even though she prefers using her hands for practice and exercises:

\begin{quote}
    ``...Having more input options allows me to choose the way that I feel most comfortable with based on my body conditions. For example, although I can only move my elbow to move my hand to tap on touch screens, which limits the input I can do with my hands, I still want to use my hand for practice because my doctor told me to practice my hands often. That is why I need more than one input method for tasks...''
\end{quote}

Furthermore, we found that our participants also like to use multiple input methods to \textbf{cope with inaccessible environments}. This indicates that having more input methods could allow people to access technology or interact with their devices in a less accessible environment. P11 further commented:

\begin{quote}
    ``...It takes effort to modify the home layout, I sometimes stay at my parents' or my sister's houses. Some of their bathrooms or stairs are not that accessible to me. A good system should have more than one input method that allows me to interact with it under complex environments...''
\end{quote}

\underline{\textit{\textbf{Appropriateness:}}} by analyzing participants' responses, we found that having multiple input modalities \textbf{reduces unwanted attention during social interactions} (P3, P4). P4 commented on her feedback of using multiple input methods to control a robot to help her eat:

\begin{quote}
    ``...Having multiple input methods allow me to prevent awkward times, such as I do not want to talk with my devices while I am having dinner with someone...''
\end{quote}

In addition to reducing unwanted attention, we also learned that having redundant input modalities could potentially \textbf{increase self-confidence} of people with upper-body motor impairments in social interactions. P10 commented on this:

\begin{quote}
    ``...Having multiple ways of controlling my devices is really cool! I could show off to my friends not only I can finish tasks independently, but also I could interact with my devices `intelligently'...''
\end{quote}

\subsubsection{Multi-modality for Input Variability}
\label{Varied Inputs throughout Task Completion}
During the interview, we also asked our participants about the perceived benefits and concerns of having multiple inputs through task completion, which means combining different input modalities to accomplish various sub-tasks of a complete task. We uncovered three main factors and the usefulness of having varied inputs throughout task completion (i.e., usability, efficiency, personalization). 

\underline{\textit{\textbf{Usability:}}} using multiple input modalities for sub-tasks enables participants to have a \textbf{natural mapping between input modalities and the actual meaning of the interface for complementary purposes} (P4, P5). P4 explained why she likes to use varied inputs for different sub-tasks:

\begin{quote}
    ``...Voice control is really bad with placing commands which require multiple steps. Therefore, all I want to control my Roomba in the future is just saying `start cleaning Roomba,' and if I need to select specific places to clean, I would just either use touch or eye-based input to make a more precise command...''
\end{quote}

Our participants mentioned that they prefer having multiple input modalities for sub-tasks to \textbf{prevent false activation and provide a more effective activation approach}. P3 further elaborated on why she prefers to use hand waving as an activation approach for voice-based systems:

\begin{quote}
    ``...If I have a smart bathroom with all the input methods you just showed, I would like to use voice because I can control it remotely. However, I would like to use hand waving or eye-based gestures as an activation approach, because I found that the current voice input accuracy and precision are not sufficient, it might falsely activate through conversations...''
\end{quote}

Having multiple input modalities to assist people with upper-body motor impairments for sub-tasks allows \textbf{additional input dimensions} for various tasks, because certain input modality does not have sufficient input dimensions, or people with upper-body motor impairments are restricted to using a subset of them (P1, P2, P3, P7, P12). P8 commented on his limited options in gaming:

\begin{quote}
    ``...I spend about ten hours in front of my computer, gaming becomes an important leisure activity for me. However, I can only play chess by using the light pointer controlled by my head. It would be great if I could have more input methods that could support me to play RPG or FPS games...''
\end{quote}

\underline{\textit{\textbf{Efficiency:}}} besides providing additional input dimensions, we learned that using multiple inputs could \textbf{reduce the effort on each single input modality}, thus reducing the complexity of learning efforts. P8 further commented on this:

\begin{quote}
    ``...I do not have full control of my fingers, I can definitely use touch to complete tasks, like document editing, but it would be really slow and can easily make me feel tired. Especially for complex tasks, that is why I prefer combining eye, head, touch, and brain control to form a hybrid control system...''
\end{quote}

By reducing the effort of a single input modality, we further learned that it would make people with upper-body motor impairments \textbf{feel more comfortable} (P5, P9). P9 elaborated on the reason for extra comfort:

\begin{quote}
    ``...Let me use gaming as an example, I have to hold a certain body posture while playing the game just by using the touch input. Having more input varieties allow me to relax and change postures during gaming...''
\end{quote}

Beyond reducing the effort of a single input modality, P10 and P11 also mentioned the benefits of \textbf{reducing the time cost} by combining multiple input modalities. P10 further explained:

\begin{quote}
    ``...Having multiple input methods on small tasks could allow me to accomplish the task faster than just using a single input. For example, I can use head or eye gestures to control where I want to shoot and then use a touch button to shoot. It would be a lot faster than I use a joystick to aim and shoot...''
\end{quote}

\underline{\textit{\textbf{Privacy and Security:}}} we mentioned in Section \ref{Comparison between Each Input Modality in ADLs} that our participants have privacy and confidentiality concerns about using existing methods in ADLs (e.g., shopping). Our participants mentioned the use of additional input for \textbf{verification and security purposes}. P1 commented on combining biometric information with a joystick for security purposes:

\begin{quote}
    ``...My current power wheelchair does not have a good authentication process, which means anyone can easily drive it by using the built-in joystick. That is why I prefer future power wheelchairs to have biometric verification for ease of use and security...''
\end{quote}

Finally, we also learned that having multiple input modalities for different purposes \textbf{ensures and extends the independence} in social activities. P10 commented on this:

\begin{quote}
    ``...A combination of multiple input methods for different small tasks could help me to become more independent while moving outside. For example, using a joystick allows me to move my wheelchair, using biometric information allows me to lock my door, and using voice can allow me to answer phone calls while my hand is occupied by controlling the joysticks. This allows me to have a complete trip independently...''
\end{quote}

\subsection{Summary of Individual versus Multimodal Input Approaches}
In this section, we presented specific reasons for choosing individual input modality (Section \ref{Comparison between Each Input Modality in ADLs}) and multimodal input alternatives (Section \ref{multimodalbenefits}) for helping people with upper-body motor impairments in ADLs. By comparing both reasons for individual and multimodal input alternatives, we found there exist similarities, such as design for reliability, reachability, and efficiency. The use of multimodal input could further extend these individual input preferences either through input redundancy (e.g., reliability, reachability) or variability (e.g., efficiency, privacy). On the other hand, we also found that multimodal input also brings opportunities that complement individual input modality. For instance, multimodal input could enhance self-confidence in social interactions for input redundancy and prevent false activation through input variability. Overall, we present how multimodal inputs provide opportunities through upper-extremity mobility limitations and adapting the complexity of contexts (Section \ref{multimodalbenefits}). We should also consider existing concerns that might also be applicable to certain multimodal inputs, such as financial burden and additional effort of installation and mapping (Section \ref{practiceperceptionchallenges}).

\section{Discussion}
\label{Discussion}
In the previous section, we presented the overall practices and challenges of different approaches in ADLs (Section \ref{practiceperceptionchallenges}), preferred applications of individual input modality and multimodal input (Section \ref{findingapplications}), and associated reasons for choosing certain individual input modality and multimodal input alternatives (Section \ref{findingimplications}).
Reflecting on these findings, we will discuss future research opportunities and design recommendations for more accessible applications of multimodal input systems. 

\subsection{Multimodal Input Towards Collaborative Experiences in ADLs}
In our findings, we showed that our participants already adopted the experiences of co-accomplishing certain ADLs with their PCAs. For instance, P11 mentioned that he used the electric toothbrush by himself, but needed a PCA to put toothpaste on the brush and press the on/off button (Section \ref{currentpractice}). Understanding the preferences for multimodal input may reduce the effort and time cost of PCAs for assisting people with upper-body motor impairments through ADLs. Furthermore, our findings of the needs and preferences of collaboratively accomplishing ADLs may further bring more opportunities for enhancing collaborative experiences \cite{branham2015collaborative} through multimodal input systems. For example, five of our participants expressed the desire to contribute to housework when their family members are busy (Section \ref{perceptionschallenges}), such as preparing meat and vegetables before their family members come back home and cook (P8). Thus, our paper proposes the following questions for future research to consider while leveraging multimodal input to support the collaborative experiences in ADLs: 1) how to support communications and interactions between PCAs and multimodal input systems? 2) how to set up collaborative tasks based on upper-extremity capabilities and different ADLs? 3) how to enable certain multimodal input systems to adapt to different collaborative tasks among ADLs with PCAs?

\subsection{Differentiate Computing Devices with Sensing Systems for Traditional ADLs}
From the findings, we recognized the disparity between technology adoption and reliance on PCAs across computing-based activities (e.g., managing finances) and traditional ADLs (e.g., toileting). Prior research uncovered the practices of multimodal input on computing devices (e.g., gaming context) \cite{wentzel2022understanding} and design considerations of sensing systems \cite{kane2020sense}. The findings from our paper uncovered the demand for taking consideration of consequences and context of use while developing multimodal input systems (e.g., safety, privacy). For example, input systems for computers would require more modifications to be able to adapt to the environment in the bathroom or for cooking scenarios \cite{li2021non}, such as location, water resistant, and surrounding objects. 

From our results, we showed that our participants prefer having various input options for different ADLs (Section \ref{perceptionschallenges}). This would likely lead to increased effort to install sensing units for various devices when used in the home context \cite{kane2020sense}. However, home environments may have unique layouts, spaces, and requirements for installment, which require designers to provide more customized solutions. Furthermore, participants expressed concerns regarding the maintenance and replacement of assistive technologies. Therefore, to reduce the effort required to exchange or repair devices and systems, it is important to explore ultra-low power or self-sustaining solutions (e.g., \cite{waghmare2020ubiquitouch}).

\subsection{Consider Actuation and Human-Robot Interaction (HRI) during Multimodal Interactions}
We also found that the significance of input modality can depend on actuation techniques when it comes to physical systems. In the context of ADLs, actuation techniques \cite{zinn2004new} include a wide spectrum of mechanisms that change the physical world, ranging from interacting with the environment, automatic doors, to connected appliances and service robots \cite{yousef2001assessment}. Working together with input modalities, these actuation techniques close the loop of interaction paradigms, examples of which can be found in P12’s case of the voice-controlled shower system, P10's robotic arm, P6's automatic door, P8's omnibot for eating and P4's Roomba in our study (Table \ref{table:ADLPractice}). Based on our findings on challenges of existing practices (Section \ref{perceptionschallenges}), leveraging actuation and robots could further support independence and reduce concerns of privacy and financial efforts. Especially, having robots could potentially benefit traditional ADLs (e.g., toileting, showering, and cooking) and reduce the reliance on upper-extremity mobility. Moreover, existing HRI research also explored how users could leverage multimodal input to interact with robots for complex tasks \cite{ajaykumarmultimodal,mohseni2019simultaneous,perzanowski2001building}. 

We propose the following research directions for HRI and supporting people with upper-body motor impairments in ADLs. First, existing research in HRI has explored how to support robot learning from multiple modalities \cite{mohseni2019simultaneous}. Future research could further combine our findings of preferred input modalities with robot learning research to support robot learning from customized multiple modalities by people with upper-body motor impairments. Second, end-user robot programming is also important through the interaction with robots, which enables end-users to overcome the complexities of specifying robot motions \cite{ajaykumarmultimodal}. The support of end-user robot programming for people with upper-body motor impairments would reduce the effort for users with upper-body motor impairments and further support the robot for complex ADLs. Third, we showed the existing high reliance on PCAs for ADLs in general. Extending from robot learning through multiple modalities \cite{mohseni2019simultaneous}, supporting active learning of robots on PCAs would strongly reduce the effort from end-users. Finally, we mentioned that some of our participants prefer working collaboratively on ADLs with PCAs or their family members. Thus, robots may become a medium to support the collaborative experiences between people with upper-body motor impairments in social interactions.

\subsection{Personalize Multimodal Designs Based on Socially Categorized ADLs}
We learned that people with upper-body motor impairments have different preferences for input modalities depending on the social context in which they will be used (Section \ref{Comparison between Each Input Modality in ADLs}), which correlates to prior research which proposed designing assistive technologies for social interactions \cite{shinohara2011shadow,li2022feels}. For example, our participants mentioned the necessity of combining biometric input with joystick input for security purposes in public. Thus, future designers may additionally consider categorizing daily tasks with more social dimensions. Based on our findings, we offer three examples of sub-categories that may be considered based on the relative social expectations in each. One might be group tasks by location and consider which inputs are more appropriate depending on whether a location is private (e.g., a bathroom) or public (e.g., a restaurant). A second category might be based on the relative expectations of particular events or activities. For example, bathing is a private activity that has relatively low expectations of social interactions, while shopping is often a completely public activity with high expectations of social connections (albeit direct or indirect contact). The third dimension we offer is based on the sensitivity or expectation of privacy for an individual in a given scenario. For this example, one might consider how a wheelchair user may again feel that the expectation of privacy in highly social activities like shopping is conversely low, whereas managing one's finances, no matter the environment or occasion, carries high expectations or desire for privacy. We offer these few examples as a starting point for future researchers and developers to consider when designing multimodal systems for use in different contexts.

\section{Limitation and Future Work}
In our study, we chose to interview people with upper-body motor impairments to understand all three research questions (RQ1 - RQ3). Although we are confident about our current contributions to the HCI and Accessibility community, there might be more opportunities by conducting contextual inquiries to further explore the detailed interactions in-depth. Beyond showing participants with different input modalities, future research could leverage technology probes to actually deploy in the living environments \cite{li2019fmt} of people with upper-body motor impairments to uncover more specific designs of multimodal input systems for ADLs.


\section{Conclusion}

In this paper, we describe the results of an interview study involving 12 people with upper-body motor impairments, which aimed to understand their current and potential future use of emerging input techniques for ADLs. We highlight the significance of incorporating new input modalities to potentially decrease reliance on PCAs and increase opportunities for independence. We assert that by understanding these opportunities based on the social- and task-based preferences of people with upper-body motor impairments, future research and development efforts can better utilize different input modalities in ADLs (Section \ref{findingimplications}).
Overall, we believe our findings contribute opportunities to support end-users' ability to choose how technology can best adapt to their unique preferences, abilities, and goals for independent and collaborative achievements of ADLs.

\bibliographystyle{ACM-Reference-Format}
\bibliography{sample-base}


\end{document}